\documentclass[12pt]{article}

\def\beq#1{\begin{equation}\label{#1}}
\def\eeq{\end{equation}}
\def\dys{\displaystyle}
\def\mst{\mathstrut}
\renewcommand{\theequation}{\arabic{section}.\arabic{equation}}
\catcode`\@=11 \@addtoreset{equation}{section}\catcode`\@=12

\newcommand{\bear}[1]{\begin{eqnarray}\label{#1}}
\newcommand{\ear}{\end{eqnarray}}

\newcommand{\N}{{\bf N}}
\newcommand{\R}{{\bf R}}
\newcommand{\fnm}{\footnotemark}
\newcommand{\fnt}{\footnotetext}
\newcommand{\sign}{ \mbox{\rm sign} }

\newcommand{\sh}{\sinh}
\newcommand{\ch}{\cosh}

\newcommand{\nn}{\nonumber}

\begin{document}

\begin{center}
\large\bf

ON SPHERICALLY SYMMETRIC  SOLUTIONS WITH HORIZON \\
IN MODEL WITH  MULTICOMPONENT ANISOTROPIC FLUID

\vspace{15pt}

\normalsize\bf

 H. Dehnen \fnm[1]\fnt[1]{Heinz.Dehnen@uni-konstanz.de},

\it Universit$\ddot{a}$t Konstanz, Fakult$\ddot{a}$t f$\ddot{u}$r Physik,
Fach  M 568, D-78457, Konstanz \\

\vspace{5pt}

\bf  V.D. Ivashchuk\fnm[2]\fnt[2]{ivas@rgs.phys.msu.su}

\it Center for Gravitation and Fundamental Metrology,
VNIIMS, 3/1 M. Ulyanovoy Str.,
Moscow 119313, Russia  and\\
Institute of Gravitation and Cosmology, PFUR,
6 Miklukho-Maklaya Str., \\ Moscow 117198, Russia

\end{center}
\vspace{15pt}

\begin{abstract}

A family of spherically symmetric solutions in the model with
 $m$-component multicomponent anisotropic fluid is considered.  
The metric
of the solution depends on  parameters $q_s > 0$, $s = 1, \ldots,m$,
relating  radial pressures and  the densities
and contains  $(n -1)m$ parameters corresponding to
Ricci-flat ``internal space'' metrics and 
obeying certain $m(m-1)/2$ (``orthogonality'') relations.
For  $q_s = 1$  (for all $s$) and certain
equations of state ($p_i^s = \pm \rho^s$) the metric  coincides with
the metric of intersecting black brane solution in the model with
antisymmetric forms.
A family of  solutions with (regular) horizon corresponding to  natural
numbers $q_s = 1,2, \ldots$ is singled out.
Certain  examples of  ``generalized simulation'' of 
intersecting $M$-branes in $D= 11$ supergravity
are considered.
The post-Newtonian parameters $\beta$ and $\gamma$ corresponding to
the 4-dimensional section of the metric are calculated.

\end{abstract}

\vspace{15pt}

PACS number(s): 04.20.Jb, 04.50.+h, 04.70.Bw


\pagebreak

\section{Introduction}

This paper is devoted 
to spherically-symmetric solutions with a horizon
in the multidimensional model with multicomponent
anisotropic  fluid defined on product
manifolds $\R \times M_0 \times \ldots \times M_n$. 
These solutions in certain cases may simulate  black brane solutions
\cite{DIM,IMS,IMS2}
(for a review on $p$-brane solutions
see \cite{IMtop} and references therein). 

We remind that $p$-brane solutions (e.g. black brane ones)
usually appear in the models with antisymmetric forms and
scalar fields (see also \cite{St}-\cite{IMp2}).
Cosmological and spherically symmetric solutions with $p$-branes are usually
obtained by the reduction of the field equations to the Lagrange equations
corresponding to Toda-like  systems \cite{IMJ}. An analogous
reduction for the models with multicomponent  "perfect" fluid was
done earlier in \cite{IM5,GIM}.

For cosmological models with
antisymmetric forms without scalar fields any $p$-brane is
equivalent to  an multicomponent anisotropic perfect fluid with the equations
of state:
  \beq{0}
    p_i = -   \rho,  \quad {\rm or} \quad  p_i =  \rho,
  \eeq

when the manifold $M_i$ belongs or does not belong to the
brane worldvolume, respectively (here $p_i$ is the  pressure
in $M_i$  and $\rho$ is the density, see Section 2).

In  this paper we find a  new family of exact
spherically-symmetric solutions
in the model with $m$-component anisotropic fluid  for 
the following equations of state
(see Appendix for more familiar form of
eqs. of state):
 \beq{0.1}
   p_r^s = - \rho^s (2q_s-1)^{-1}, \qquad p_0^s = \rho^s (2q_s-1)^{-1},
 \eeq
and
 \beq{0.2}
   p_i^s= \left(1-\frac{2U^s_i}{d_i} \right) \rho^s/(2q_s-1),
 \eeq
 $i > 1$, $s = 1, \ldots, m$,
where for $s$-th component: 
 $\rho^s$ is a density, $p_r^s$ is a radial pressure,
 $p_i^s$ is a pressure in $M_i$, $i = 2, \dots, n$.
Here parameters $U^s_i$ ($i > 1$)  and
the parameters $q_s = U^s_1 > 0$ obey 
the following 
``orthogonality'' relations (see also  Section 2 below)
  \beq{0.3}
     B_{sl} = 0, \qquad s \neq l
   \eeq
where
  \beq{0.3a}
    B_{sl} \equiv  \sum_{i=1}^{n} \frac{U^{s}_i U^l_i}{d_i}
     + \frac{1}{2-D} \left(\sum_{i=1}^{n } U^s_i \right)
                      \left(\sum_{j=1}^{n } U^l_j \right),
   \eeq
$q_s \neq 1/2$; and  $s,l = 1, \ldots, m$.
The manifold $M_0$ is $d_0$-dimensional sphere
in our case and $p_0^s$ is the pressure in the tangent direction.

The one-component case was considered earlier in \cite{DIM}.  
For special case with $q_s = 1$ see \cite{IMS}
and \cite{IMS2} (for one-component and multicomponent case,
respectively).

The paper is organized as follows. In Section 2
the model with multicomponent (anisotropic or ``perfect'')
fluid is formulated. In Section 3 a subclass
of spherically symmetric solutions 
(generalizing solutions from \cite{IMS2}) is presented
and  solutions with (regular) horizon 
corresponding to integer $q_s$ are singled out.
Section 4  deals with certain examples of 
two-component solutions
in dimension $D=11$ 
containing for $q_s =1$ intersecting
$M 2 \cap M2$,  $M2 \cap M 5$ and $M5 \cap M5$ black brane metrics.
In Section 5  the post-Newtonian
parameters for the 4-dimensional section
of the metric are calculated. In the Appendix
a class of general spherically symmetric solutions
in the model under consideration is presented.

\section{The model}

Here, we consider a family of spherically symmetric   solutions to Einstein 
equations with an multicomponent anisotropic fluid matter source
  \beq{1.1} 
   R^M_N - \frac{1}{2}\delta^M_N R = k T^M_N 
  \eeq
defined on the manifold
  \beq{1.2}
    \begin{array}{rlll}
    M = &{\R}_{.} \quad \times &(M_{0}=S^{d_0})\quad \times &(M_1 =
    {\R}) \times M_2 \times \ldots \times M_n,\\ &^{radial
    \phantom{p}}_{variable}&\quad^{spherical}_{variables}&\quad^{time}
   \end{array}
   \eeq 
with the block-diagonal metrics
  \beq{1.2a} 
     ds^2=
     e^{2\gamma (u)} du^{2}+\sum^{n}_{i=0} e^{2X^i(u)}
     h^{(i)}_{m_in_i} dy^{m_i}dy^{n_i }. 
   \eeq

Here $\R_{.} = (a,b)$ is interval. The manifold $M_i$ with the metric
$h^{(i)}$, $i=1,2,\ldots,n$, is the Ricci-flat space of dimension
$d_{i}$:

   \beq{1.3} 
    R_{m_{i}n_{i}}[h^{(i)}]=0, 
   \eeq and $h^{(0)}$ is
standard metric on the unit sphere $S^{d_0}$

   \beq{1.4} 
     R_{m_{0}n_{0}}[h^{(0)}]=(d_0-1)h^{(0)}_{m_{0}n_{0}},
   \eeq 
  $u$ is radial variable, $\kappa$ is the multidimensional gravitational
constant, $d_1 = 1$ and $h^{(1)} = -dt \otimes dt$.

The  energy-momentum tensor is adopted in the following form
  \beq{1.5a}
    T^{M}_{N}= \sum_{s = 1}^{m} T^{(s)M}_{N}, 
  \eeq
where
 \beq{1.5b}
   T^{(s)M}_{N}= {\rm diag}(-(2q_s-1)^{-1} \rho^s,
   (2q_s-1)^{-1} \rho^s \delta^{m_{0}}_{k_{0}}, - \rho^s , p_2^s
   \delta^{m_{2}}_{k_{2}}, \ldots ,  p_n^s \delta^{m_{n}}_{k_{n}}),
 \eeq
 $q_s > 0$ and $q_s \neq 1/2$.
The pressures $p_i^s$  and
the density $\rho^s$ obeys the relations
  (\ref{0.2}) with  constants $U_i^s$, $i >1$.

The ``conservation law'' equations
  \beq{con}
  \nabla_{M} T^{(s)M}_N = 0
  \eeq
are assumed to be valid for all $s$.
  
In what follows we put $\kappa = 1$ for simplicity.

\section{Exact solutions }

Let us  define
  \bear{2.1}
  &&1^{o}.\quad U^s_0 = 0,
      \\ \label{2.1a}
  &&2^o.\quad U^s_1 = q_s,
      \\ \label{2.1c}
  &&3^o. \quad (U^s,U^l) = U^s_i G^{ij} U^l_j 
         \ear
where $U^s = (U^s_i)$ is $(n+1)$-dimensional vector 
and
  \beq{2.2a}
     G^{ij}=\frac{\delta^{ij}}{d_i} + \frac{1}{2-D}
   \eeq
are components of the matrix inverse to the matrix of the
minisuperspace metric \cite{IM0,IMZ}
   \beq{2.2}
     (G_{ij}) = (d_i \delta_{ij} - d_i d_j), 
   \eeq
  $i,j = 0, \ldots, n$,
and  $D=1+\sum\limits_{i=0}^n {d_i}$ is the total dimension.

In our case the scalar products (\ref{2.1c}) 
 are given by relations:
   \beq{2.1d}
    (U^s,U^l) = B_{kl}
   \eeq
with $B_{kl}$ from (\ref{0.3a})  and hence due to  (\ref{0.3})
vectors $U^s$ are mutually orthogonal, i.e.
    \beq{2.1e}
        (U^s,U^l) = 0, \qquad s \neq l.
     \eeq

It is proved in Appendix that the relation
   $1^{o}$ implies 
    \beq{2.1b}
       (U^s,U^s)  > 0,
    \eeq
 for all $s$.

For the equations of state (\ref{0.1}) and (\ref{0.2})
with parameters obeying (\ref{0.3}) 
we have obtained the following spherically symmetric
solutions  to the Einstein equations (\ref{1.1})
(see Appendix)
    \bear{12}
    ds^{2} = J_{0}\left( \frac{dr^{2}}{1-\frac{2\mu}{r^{d}}} +
    r^{2} d \Omega^2_{d_0} \right)
    - J_1 \left(1-\frac{2\mu}{r^{d}}\right)dt^{2}
    \\ \nn 
     + \sum_{i=2}^{n}  J_{i} h^{(i)}_{m_{i}n_{i}} dy^{m_{i}}dy^{n_{i}},
    \\ \label{13}
    \rho^s=\dys \frac{\mst (2q_s-1)(dq_s)^2
    P_s(P_s+2\mu)(1 - 2 \mu r^{-d})^{q_s-1}}
    {2 (U^s,U^s) (\prod_{s=1}^{m} H_s)^2 J_0 r^{2d_0}},
    \ear
by methods similar to obtaining $p$-brane
solutions \cite{IMJ}. Here $d=d_0-1$, $d \Omega_{d_0}^2=
    h^{(0)}_{m_{0}n_{0}} dy^{m_{0}}dy^{n_{0}}$
is spherical element, the metric factors
 
  \bear{2.3}
   J_{i} = \prod_{s= 1}^{m} H_s^{-2U^{si}/(U^s,U^s)},
   \\ \label{2.3a} 
   H_s =    1+ \frac{P_s}{2 \mu}
   \left[1 - \left(1 - \frac{2 \mu }{r^{d}}\right)^{q_s} \right];
 \ear 
  $P >0$, $\mu >0$ are constants and

   \beq{2.4}
     U^{si} = G^{ij}U^s_{j}  = \frac{U^s_i}{d_i} + \frac{1}{2-D}
     \sum_{j=0}^{n}U^s_j.
   \eeq

Using (\ref{2.4}) and  $U_0^s =0$
we get
  \beq{2.4a}
   U^{s0} =  \frac{1}{2-D} \sum_{j=0}^{n}U^s_j
  \eeq
and hence one can rewrite (\ref{12}) as follows
  \bear{12a}
    ds^{2} = J_{0} \left[ \frac{dr^{2}}{1-\frac{2\mu}{r^{d}}}
    + r^{2} d \Omega^2_{d_0} -  
     \left(\prod_{s=1}^{m} H_s^{-2q_s/(U^s,U^s)}\right) 
    \left(1-\frac{2\mu}{r^{d}} \right)dt^2 +    \right.
    \\ \nn
    \left.
    {\qquad} + \sum\limits_{i=2}^{n}
    \left(\prod_{s= 1}^{m} H_s^{-2U^s_i/(d_i(U^s,U^s))}\right)
     h^{(i)}_{m_{i}n_{i}} dy^{m_{i}}dy^{n_{i}}
    \right].
    \ear

These solutions are special case of general solutions
spherically symmetric solutions obtained in Appendix
by method suggested in \cite{GIM}.

{\bf Black holes for natural $q_s$.}

For natural
  \beq{12n}
     q_s= 1,2, \ldots, 
  \eeq
the metric  has a  horizon at $r^d = 2 \mu = r_h^2$.
Indeed, for these values of $q_s$ the functions $H_s(r) > 0$
are smooth in the interval  $(r_{*}, + \infty)$
for some  $r_{*} < r_h$. For odd $q_s= 2m_s +1$ 
(for all $s$) one get $r_{*} = 0$.

A global structure of the black hole solution corresponding
to these values of $q_s$ will be a subject of a separate
publication.

It was shown in \cite{DIM} that in one-component case
for $2 U^{s 0} \neq -1$ and $0< q_s<1$ one get
 singularity at $r^d \to 2 \mu$.

 {\bf Remark.} For  non-integer $q_s>1$ 
the function $H_s(r)$ have a
non-analytical behavior in the vicinity
of $r^d = 2 \mu$.  In this case one may conject
that  the limit $r^d \to 2 \mu$  corresponds 
to the singularity (in general case)
but here a separate investigation is needed.

\section{Examples: generalized simulation of intersecting black branes}

The solutions with a horizon from the previous section  allow us
to simulate the intersecting black brane solutions \cite{IMtop}
in the model with antisymmetric forms without scalar fields \cite{IMS}
when all $q_s =1 $.

These solutions may be also generalized to the case of
general natural $q_s \in \N$.
In this case the parameters $U^s_i$ and the pressures have  the following
form
 \beq{3.1a}
   \begin{array}{ccccccr}
    U^s_i & = &q_s d_i , \quad & p_i^s & = & -\rho^s , &  i\in I_s;
    \\
    &  & 0 , & \quad & &(2q_s -1)^{-1} \rho^s,  &  i \notin I_s.
  \end{array}
 \eeq

Here  $I_s = \{ i_1, \ldots,  i_k \} \in \{1, \ldots n \}$
is the index set \cite{IMtop} corresponding to ``brane'' submanifold 
 $M_{i_1}  \times \ldots \times M_{i_k}$. 

The "orthogonality" relations  (\ref{0.3}) lead us to  the following 
dimension of intersection of brane submanifolds \cite{IMtop}
  \beq{3.1b}
   d(I_s \cap I_l)= \frac{d(I_s) d(I_l)}{D-2}, \qquad s \neq l,
  \eeq
where $d(I_s)= \sum_{i \in I_s} d_i $ is dimension of  $p$-brane
worldvolume. 

  {\bf Remark.} The set of Diophantus equations 
  (\ref{3.1b}) was solved
  explicitly in \cite{IM12} for so-called ``flower'' Ansatz
  from \cite{AV}. The solution in this case takes place for 
  infinite number of dimensions $D = 6,10,11,14,18,20,26,27,...$ etc.

 As an example, here  we consider a ``generalized simulation'' of
 intersecting     $M2 \cap M5$, $M2 \cap M2$
 and  $M5 \cap M5$ black branes
 in $D = 11$ supergravity. In what follows functions
  $H_s$, $s =1,2$, are defined in (\ref{2.3a}).

$a).$ For  an analog  of intersecting
    $M2 \cap M5$ branes the metric reads:

 \bear{3.3}
     ds^2 = H_{1}^{1/(3q_1)} H_{2}^{2/(3q_2)} \Biggl[\dys\frac{\mst
        dr^{2}}{1-{2\mu}/{r^{d}}} + r^{2} d \Omega^2_{d_0}
    \\ \nn   
      - H_{1}^{-1/q_1}  H_{2}^{-1/q_2}
      \biggl\{ \left(1-\frac{2\mu}{r^d}\right)
      dt^2 + dy^{m_{2}}dy^{m_{2}} \biggr\}  \Biggr. \\ \nn 
     \Biggl.  
    + H_{2}^{-1/q_2} h^{(3)}_{m_{3}n_{3}} dy^{m_{3}}dy^{n_{3}}
    + H_{1}^{-1/q_1} dy^{m_{4}}dy^{m_{4}}
    + h^{(5)}_{m_{5}n_{5}} dy^{m_{5}}dy^{n_{5}}\Biggr],
  \ear
where $M2$-brane includes three one-dimensional spaces: $M_2$,
 $M_4$ and the time manifold $M_1$; and $M5$-brane includes 
 $M_1$,$M_2$ and $M_3$ ($d_3 =4$).

 $b).$ An analog of two electrical $M2$ branes 
 intersecting on the time manifold  has the following metric
 \bear{3.3a}
    ds^2 = H_{1}^{1/(3q_1)} H_{2}^{1/(3 q_2)} \Biggl[\dys\frac{\mst
           dr^2}{1- 2\mu/r^{d}} + r^{2} d \Omega^2_{d_0} 
           \\ \nn
        - H_{1}^{-1/q_1} H_{2}^{-1/q_2} 
        \left( 1- \frac{2\mu}{r^d} \right)
        dt^2 \Biggr.
        \\ \nn    \Biggl. 
      + H_{1}^{-1/q_1} h^{(2)}_{m_2 n_2}  dy^{m_2}dy^{n_2} + 
        H_{2}^{-1/q_2}  h^{(3)}_{m_{3} n_{3}}  dy^{m_3}dy^{n_3} + 
        h^{(4)}_{m_4 n_4} dy^{m_4} dy^{n_4} \Biggr], 
  \ear
where $d_2 = d_3 = 2$.

 $c).$ For an analog of
two intersecting $M5$ branes the dimension of 
intersection is $4$ and the metric reads

\bear{3.3b}
    ds^2 = H_{1}^{2/(3 q_1)}H_{2}^{2/(3q_2)} \Biggl[\dys\frac{\mst
        dr^2 }{1- 2\mu/r} + r^2 d \Omega^2_2 
    \\ \nn 
      - {H_1}^{-1/q_1}{H_{2}}^{-1/q_2}\biggl\{
    \left(1-\frac{2\mu}{r}\right) dt^2   + h^{(2)}_{m_2 n_2}
    dy^{m_2}dy^{n_2} \biggr\} \Biggr.
     \\ \nn  \Biggl.
    + H_{1}^{-1/q_1} h^{(3)}_{m_{3} n_{3}} dy^{m_3}dy^{n_3} +
    H_{2}^{-1/q_2} h^{(4)}_{m_{4} n_{4}} dy^{m_4}dy^{n_4}\Biggr]. 
   \ear

Here $d_0 =d_3 = d_4 = 2$ and $d_2 = 3$.

For the density of $s$-th component we get
in any of these  three cases
    \beq{3.4}
     \rho^s= \frac{(2q_s-1) d^2
     P_s(P_s+2\mu)(1 - 2 \mu r^{-d})^{q_s-1}}
     {4 ( H_1 H_2)^2 J_0 r^{2d_0}},
    \eeq
where  
   \beq{3.5}
    J_0 = \prod_{s= 1}^{2} H_s^{d(I_s)/(9q_s)}
   \eeq
and $d(I_s) = 3, 6$ for  $M2, M5$ branes, respectively.

 \section{Physical parameters}

 \subsection{Gravitational mass and PPN parameters}

Here we put $d_0  =2 \ (d =1)$. Let us
consider the 4-dimensional space-time section of the metric
 (\ref{12a}). Introducing a new radial variable by the relation:

 \beq{3.7} r = R \left(1 + \frac{\mu}{2 R}\right)^2, \eeq
we rewrite the $4$-section in the following form:
 \bear{3.8}
  ds^2_{(4)} = \left(\prod_{s=1}^{m} H^{- 2 U^{s0}/(U^s,U^s)}\right) 
   \left[ - \left(\prod_{s=1}^{m} H_s^{- 2q_s/(U^s,U^s) } \right)
    \left(\frac{1-\frac{\mu}{2 R}}{1+\frac{\mu}{2 R}}\right)^2 dt^2
       \right.
        \\ \nn    \left.
    + \left( 1+\frac{\mu}{2 R} \right)^4 \delta_{ij}
   dx^i dx^j \right]
  \ear
$i,j = 1,2,3$. Here $R^2 = \delta_{ij} x^i x^j$.

The parameterized post-Newtonian (Eddington) parameters are
defined by the well-known relations

 \bear{3.9}
   g^{(4)}_{00} = - (1-2V+2\beta V^2) + O (V^3), \\
   \label{3.10} g^{(4)}_{ij} = \delta_{ij} (1 + 2 \gamma V) + O(V^2),
 \ear
  $i,j = 1,2,3$. Here
      \beq{3.11}
      V=\frac{GM}{R}
     \eeq
is
the Newtonian potential, $M$ is the gravitational mass and $G$ is
the gravitational constant.

From (\ref{3.8})-(\ref{3.10}) we obtain:
  \beq{3.12}
   GM = \mu +  \sum_{s=1}^{m}
  \frac{P_s q_s(q_s+U^{s0})}{(U^s,U^s)} 
  \eeq
and
 \bear{3.13}
  \beta - 1= \sum_{s=1}^{m}
  \frac{|A_s|}{(GM)^2} (q_s + U^{s0}), \\
   \label{3.13a}
  \gamma - 1= - \sum_{s=1}^{m} \frac{ P_s q_s}{(U^s,U^s) GM} (q_s + 2U^{s0}),
\ear
where 
  \beq{a} 
  |A_s| = \frac12 q_s^2 P_s (P_s + 2 \mu)/(U^s,U^s)  
  \eeq
(see Appendix) or, equivalently, 
  \beq{p}
   P_s = - \mu + \sqrt{\mu^2 + 2 |A_s| (U^s,U^s) q_s^{-2}} > 0.
  \eeq

For fixed $U^s_i$
the parameter $\beta -1$ is proportional to the  ratio of two 
quantities: the weighted sum of 
multicomponent anisotropic fluid density parameters $|A_s|$
and the gravitational radius squared
$(GM)^2$.

\subsection{Hawking temperature}

The Hawking temperature of the black hole may be calculated
using the well-known relation \cite{York}
  \beq{Y}
      T_H  = \frac{1}{4 \pi \sqrt{-g_{tt} g_{rr}}}
      \frac{d(-g_{tt})}{dr} \Biggl|_{_{\ horizon}}.  
  \eeq
We get
  \beq{H}
   T_H =
        \frac{d}{4 \pi (2 \mu)^{1/d}}
     \prod_{s=1}^{m}
\left(1 + \frac{P_s}{2\mu}\right)^{- q_s/(U^s,U^s)}.
\eeq
  Here all $q_s$ are natural numbers. 

For any of  $D=11$ metrics from Section 4 
the Hawking temperature reads
\\
 $T_H =
  \frac{d}{4 \pi (2 \mu)^{1/d}}
\prod_{s=1}^{2} \left(1 + \frac{P_s}{2\mu}\right)^{- 1/(2q_s)}$.
 \\

\section{Conclusions}

In this paper, using the methods developed earlier for obtaining perfect
fluid and p-brane solutions, we have  considered a family of spherically
symmetric solutions  in the model with m-component  anisotropic fluid when
the equations of state (\ref{0.1})- (\ref{0.3}) are imposed.  The
metric of any solution  contains  $(n -1)$ Ricci-flat "internal" space
metrics and depends upon a set of  parameters $U^s_i$, $i > 1$.

For $q_s = 1$ (for all $s$) and  certain equations of state 
(with $ p_i^s = \pm \rho^s$) 
the metric of the solution  coincides with that of 
intersecting black brane 
 solution in the model with antisymmetric forms without
dilatons \cite{IMS2}.
For natural numbers $q_s = 1,2, \ldots$  we have obtained
a family of solutions with regular horizon.

Here we  have considered
three examples of  solutions with horizon, that
simulate (by fluids) binary intersecting 
$M2$ and  $M5$  black branes in $D=11$ supergravity.

We  have also calculated (for  possible estimations of observable effects
of extra dimensions) the post-Newtonian parameters $\beta$ and $\gamma$
corresponding to the 4-dimensional section of the metric and the Hawking
temperature as well.  

\newpage

{\bf Acknowlegments}

This work was supported in part by the Russian Ministry of
Science and Technology, Russian Foundation for Basic Research
(RFFI-01-02-17312-a)and DFG Project (436 RUS 113/678/0-1(R)).
V.D.I. thanks colleagues from the Physical Department of
the University of Konstanz for 
hospitality during his visit in August-October 2003.


\renewcommand{\theequation}{\Alph{subsection}.\arabic{equation}}
\renewcommand{\thesection}{}
\renewcommand{\thesubsection}{\Alph{subsection}}
\setcounter{section}{0}

\section{Appendix}

\subsection{Lagrange representation}

It is more convenient for finding
of exact solutions, to write the stress-energy tensor
in cosmological-type form
 \beq{1.5} (T^{(s)M}_{N})= {\rm diag}(-{\hat{\rho^s}},{\hat p}_0^s
   \delta^{m_{0}}_{k_{0}}, {\hat p}_{1}^s
   \delta^{m_{1}}_{k_{1}},\ldots , {\hat p}_n^s
   \delta^{m_{n}}_{k_{n}}),
 \eeq
where $\hat{\rho}^s$ and ${\hat p}_i^s$
are "effective"  density and pressures
of $s$-th component, respectively, depending
upon the radial variable $u$ and
the physical density $\rho^s$
and pressures $p_i^s$ are related to the effective
("hat") ones by formulas
 \beq{1.7a} \rho^s = - {\hat p}_1^s, 
    \quad p_r^s = - \hat{\rho}^s, \quad
     p_i^s = \hat{p}_i^s, \quad (i \neq 1),
 \eeq
$s = 1, \ldots, m$.

The equations of state may be written in the following
form
 \beq{1.7}
    {\hat p}_i=\left(1-\frac{2U^s_i}{d_i}\right){\hat{\rho^s}},
 \eeq
where $U^s_i$ are constants, $i= 0,1, \ldots, n$.
It follows from (\ref{1.7a}), (\ref{1.7}) and $U^s_1 = q_s$
that
  \beq{1.7b}
      \rho^s = (2q_s - 1) \hat{\rho^s}.
  \eeq

The ``conservation law'' equations
$\nabla_{M} T^{(s) M}_N = 0$ may be written,
due to relations (\ref{1.2a}) and (\ref{1.5}) in the following form:
   \beq{5.7} 
   \dot{\hat{\rho^s}} +\sum_{i=0}^n
     d_i\dot{X^i}({\hat{\rho^s}} +{\hat p}_i^s )=0. 
   \eeq
Using the equation of state (\ref{1.7}) we get
  \beq{5.7a}
   \hat{\rho}^s = - A_s e^{2U^s_i X^i -2 \gamma_0 },
  \eeq
where $\gamma_0(X)= \sum\limits_{i=0}^{n} d_{i}X^{i}$ and $A_s$
are constants.

The Einstein equations (\ref{1.1})  with the relations
(\ref{1.7}) and (\ref{5.7a}) imposed are equivalent to the
Lagrange equations for the Lagrangian
  \beq{lag} L = \frac{1}{2}
   e^{-\gamma+\gamma_0(X)}G_{ij}\dot{X}^{i}\dot{X}^{j}
     -e^{\gamma-\gamma_0(X)} V, 
  \eeq
where
   \beq{5.32n}
    V= \frac{1}{2} d_0 (d_0 -1) \exp(2U^0_i X^i) +
       A_s \exp(2 U^s_i X^i)
   \eeq
is the potential and the components
of the minisupermetric $G_{ij}$ are defined in (\ref{2.2}),
  \beq{5.8}
    U^{0}_i X^i = -X^0 + \gamma_0(X), \qquad U^{0}_i  =- \delta^0_i + d_i,
  \eeq
 $i = 0, \ldots, n$ (for cosmological case see \cite{IM5,GIM}).

For $\gamma=\gamma_0(X)$, i.e. when the harmonic time gauge
is considered, we get the set of Lagrange equations
for the Lagrangian
  \beq{5.31n}
    L=\frac12G_{ij} \dot X^i \dot X^j-V,
  \eeq
with the zero-energy constraint imposed
  \beq{5.33n}
   E=\frac12G_{ij} \dot X^i \dot X^j + V =0.
  \eeq

It follows from the restriction
   $U_0^s = 0$ that
  \beq{5.43a}
    (U^0,U^s)  \equiv U^0_i G^{ij}U_j^s = 0.
  \eeq

Indeed, the contravariant components $U^{0i}=G^{ij} U^0_j$ 
are the following ones
\beq{5.43b}
U^{0i}=-\frac{\delta_0^i}{d_0}.
\eeq

Then we get $(U^0,U^s)  = U^{0i} U_i^s = - U_0^s/d_0 =0$.
In what follows we also use the formula
\beq{5.43c}
(U^0,U^0)   = \frac{1}{d_0} - 1 < 0
\eeq
for $d_0 >1$.

Now we prove that $(U^s,U^s) > 0$ for all $s >0$. Indeed, minisupermetric
has the signature $(-,+,\ldots,+)$ \cite{IM0,IMZ}, vector $U^0$
is time-like and orthogonal to any vector $U^s \neq 0$. Hence
any vector $U^s$ is space-like.

\addtocounter{section}{1} \setcounter{equation}{0}

\subsection{General spherically symmetric solutions}

When the orthogonality relations  (\ref{5.43a}) and
(\ref{2.1e}) are satisfied the Euler-Lagrange equations for the
Lagrangian (\ref{5.31n}) with the potential (\ref{5.32n}) have the
following solutions (see relations from \cite{GIM} adopted for our
case):

  \beq{5.34n}
    X^i(u)= -
     \sum_{\alpha=0}^{m}
     \frac{U^{\alpha i}}{(U^{\alpha},U^{\alpha})}
     \ln|f_{\alpha}(u-u_{\alpha})| + c^i u + \bar{c}^i, 
  \eeq
where
  $u_{\alpha}$
are integration constants; and vectors $c=(c^i)$ and 
$\bar c=(\bar c^i)$ are dually-orthogonal 
to co-vectors 
 $U^{\alpha }=(U^{\alpha }_i)$, 
i.e. they satisfy the linear constraint relations
  \bear{5.47n}
   U^0(c)= U^0_i c^i = -c^0+\sum_{j=0}^n d_j c^j=0, \\
   \label{5.48n} U^0(\bar c)= U^0_i \bar c^i = 
    -\bar c^0+\sum_{j=0}^n d_j \bar c^j=0, \\
   \label{5.49n} U^s(c)= U^s_i c^i=0,\\
   \label{5.50n} U^s(\bar c)=  U^s_i \bar c^i=0. 
  \ear
 Here
  \beq{A.7}
    \begin{array}{rlll}
     f_{\alpha}(\tau)=
     & R_{\alpha} \dys\frac{\mst
    \sh(\dys\sqrt{\mst C_{\alpha}}\tau)}{\dys\sqrt{\mst C_{\alpha}}},
    & C_{\alpha} > 0,&
    \eta_{\alpha} = +1 ,
    \\  & R_{\alpha} \dys\frac{\mst\ch(\dys\sqrt{\mst
    C_{\alpha}}\tau)}{\dys\sqrt{\mst C_{\alpha}}},& C_{\alpha}>0,
        & \eta_{\alpha} = -1 ,
    \\  & R_{\alpha} \dys\frac{\mst\sin(\dys\sqrt{\mst
   |C_{\alpha}|}\tau)}{\dys\sqrt{\mst |C_{\alpha}|}},& C_{\alpha}<0,
   & \eta_{\alpha} = + 1 , \\\\
   & R_{\alpha} \tau,& C_{\alpha}=0,&
  \eta_{\alpha} = + 1 ,
  \end{array}
  \eeq
$\alpha = 0, \ldots, m$; where $R_0 = d_0-1$, $\eta_0 = 1$,
$R_s =\sqrt{2|A_s|(U^s,U^s)}$, $\eta_s = - \sign A_s$
($s = 1, \ldots, m$).

The zero-energy constraint, corresponding to the solution (\ref{5.34n})
reads

  \beq{A.17}
    E = \frac12 \sum_{\alpha = 0}^m
    \frac{C_{\alpha}}{(U^{\alpha} , U^{\alpha})} + \frac12
     G_{ij}c^ic^j=0 . 
  \eeq

{\bf Special solutions.}
The (weak) horizon condition (i.e. infinite
time of propagation of light for $u \to +\infty$) lead us to the
following integration constants
   \bear{5.67}
     \bar{c}^i  & = &  0,\\
      c^i & = & \bar{\mu}  \sum_{\alpha = 0}^m
    \frac{U^{\alpha}_1 U^{\alpha i}}{(U^{\alpha},U^{\alpha})} -
    \bar{\mu} \delta^i_1, \\
      \label{5.68}
     C_{\alpha}  & = & (U^{\alpha}_1)^2  \bar{\mu}^2,
   \ear
where $\bar{\mu} > 0$, $\alpha=0, \ldots, m $.
For analogous choice of parameters in  
$p$-brane case see \cite{BIM,IMJ,IMtop}.

We also introduce a new radial variable $r = r(u)$ by relations

\beq{5.69}
\exp( - 2\bar{\mu} u) = 1 - \frac{2\mu}{r^d},  \quad
\mu = \bar{\mu}/d >0, \quad  d = d_0 -1,
\eeq
and put $u_s < 0$ and  $A_s  < 0 $ for all $s$ and also  $u_0 = 0$.

The relations of the Appendix imply the formulae
(\ref{12}) and (\ref{13})
for the solution from Section 3 with
\beq{5.70}
H_s =  \exp(- \bar{\mu} q_s u) f_s(u - u_s), \qquad
A_s = - \frac{(dq_s)^2}{2(U^s,U^s)} P_s ( P_s +2 \mu),
\eeq
$P_s > 0$.


\newpage

\small
\begin{thebibliography}{99}


\bibitem{DIM}
H. Dehnen, V.D. Ivashchuk and  V.N. Melnikov,
On black hole solutions in model
with anisotropic fluid,  gr-qc/0211049,
to appear in Grav. Cosmol.

\bibitem{IMS}
V.D. Ivashchuk, V.N. Melnikov and A.B. Selivanov,
Multidimensional black hole solutions in model
with anisotropic fluid,  {\it Grav. Cosmol.} {\bf 7}, 4(12),
308-310 (2001); gr-qc/0205103

\bibitem{IMS2}
V.D. Ivashchuk, V.N. Melnikov and A.B. Selivanov,
Simulation of intersecting black brane 
solutions,  hep-th/0211247, to appear in Grav. Cosmol.

\bibitem{IMtop}
V.D. Ivashchuk and V.N. Melnikov, Exact solutions in
multidimensional gravity with antisymmetric forms, topical review,
{\it Class. Quantum Grav.},  {\bf 18 } (2001) R87-R152;
hep-th/0110274.

\bibitem{St}
K.S. Stelle, "Lectures on supergravity p-branes ",
hep-th/9701088.

\bibitem{HS}
G.T. Horowitz and A. Strominger, Black string and p-branes, {\it
Nucl. Phys.} {\bf B 360}, 197-209 (1991)

\bibitem{DL}
M.J. Duff and J.X. Lu, {\it Nucl. Phys.} {\bf B 354}, 129
(1991).

\bibitem{LuP}
H. L\"u and C.N. Pope, Black p-branes and their vertical
dimensional reduction, {\it Nucl. Phys.} {\bf B 489}, 264-278
(1997); hep-th/9609126.

\bibitem{CT}
M. Cvetic and A. Tseytlin, Non-extreme black holes from
non-extreme intersecting $M$-branes, {\it Nucl. Phys.} {\bf B
478}, 181-198 (1996); hep-th/9606033.

\bibitem{C}
M. Costa, Black composite M-branes,{\it Nucl. Phys.} {\bf B
490}, 603-614 (1997); hep-th/9610138.

\bibitem{AIV}
I.Ya. Aref'eva, M.G. Ivanov and I.V. Volovich, Non-extremal intersecting
p-branes in various dimensions, {\it Phys. Lett.} {\bf B 406}, 44-48
(1997); hep-th/9702079.

\bibitem{Oh}
N. Ohta, Intersection rules for non-extreme p-branes, {\it Phys. Lett.}
{\bf B 403}, 218-224 (1997); hep-th/9702164.

\bibitem{BIM}
K.A. Bronnikov, V.D. Ivashchuk and V.N. Melnikov, The
Reissner-Nordstr\"om problem for intersecting electric and
magnetic  p-branes,  {\it Grav. Cosmol.} {\bf 3},  3(11), 203-212
(1997); gr-qc/9710054.

\bibitem{IMJ}
V.D. Ivashchuk and V.N. Melnikov, Multidimensional classical and
quantum cosmology with intersecting $p$-branes, {\it J. Math.
Phys.}, {\bf 39}, 2866-2889 (1998); hep-th/9708157.

\bibitem{IMp2}
V.D. Ivashchuk and V.N. Melnikov, Black hole p-brane solutions for
general intersection rules, {\it Grav. Cosmol.}, {\bf 6}, 1(21),
27-40 (2000); hep-th/9910041.

\bibitem{IM0}
V.D. Ivashchuk and V.N. Melnikov, Perfect-fluid type  solution  in
multidimensional cosmology, {\it Phys. Lett.}  {\bf A 136},
465-467 (1989).

\bibitem{IMZ}
V.D. Ivashchuk,  V.N. Melnikov and A.I. Zhuk,
On Wheeler-DeWitt equation in multidimensional cosmology,
{\it Nuovo Cimento } {\bf B 104}, 5, 575-581  (1989).

\bibitem{IM5}
V.D. Ivashchuk and V.N. Melnikov, Multidimensional cosmology with
$m$-component perfect fluid, {\it Int. J. Mod. Phys.} {\bf D 3},
4, 795-811 (1994); gr-qc/ 9403063.

\bibitem{GIM}
V.R. Gavrilov, V.D. Ivashchuk and V.N. Melnikov, Integrable
pseudo-Euclidean Toda-like systems in multidimensional cosmology
with multicomponent perfect fluid, {\it J. Math. Phys } {\bf 36},
5829-5847 (1995).

\bibitem{IM12}
 V.D. Ivashchuk and V.N. Melnikov,
 Intersecting p-brane solutions in
 Multidimensional gravity and M-Theory,
 hep-th/9612089; {\it Grav. Cosmol.},
 {\bf 2}, No. 4 (8),  297-305 (1996).

\bibitem{AV}
I.Ya. Aref'eva and  A.I. Volovich, 
Composite p-branes in diverse-dimensions, 
{\it Class. Quantum Grav.} {\bf 14}, 2991-3000
(1997); hep-th/9611026.

\bibitem{York}
J.W. York, {\it Phys. Rev.} {\bf D 31}, 775 (1985).

\end{thebibliography}
\end{document}